\documentclass{caosp305}

\usepackage{graphicx}
\usepackage{subcaption}
\usepackage{amssymb}

\usepackage{natbib}
\articleNo{123}
\pubyear{2017}
\volume{35}
\volnumber{3}
\firstpage{1}
\received{Dec. 7, 2017}
\accepted{Dec. 7, 2017}

\def\BibTeX{{\rm B\kern-.05em{\sc i\kern-.025em b}\kern-.08em
             T\kern-.1667em\lower.7ex\hbox{E}\kern-.125emX}}

\begin{document}

\hauthor{J.\,Sikora, G.\,Wade, and J.\,Power}

\title{Understanding the fossil magnetic fields of Ap/Bp stars}
\subtitle{Conclusions from a volume-limited survey}

\author{
        J.\,Sikora \inst{1,} \inst{2}
      \and 
        G.\,Wade \inst{2}
      \and 
        J.\,Power \inst{1,} \inst{2}
       }

\institute{
           Dept. of Physics, Engineering Physics \& Astronomy, Queen's University\\
           Kingston, ON, Canada,
           \email{james.sikora@queensu.ca}
         \and
           Dept. of Physics and Space Science, Royal Military College of Canada\\
           Kingston, ON, Canada
          }

\date{Dec. 7, 2017}

\maketitle

\begin{abstract}
Various observational properties of Ap/Bp stars have been well-established such as the often-cited 
$10\,\%$ incidence rate of strong, organized magnetic fields amongst all A- and B-type stars. However, 
these inferences have generally been drawn from surveys biased towards the strongest most easily 
detectable fields. A volume-limited spectropolarimetric survey of all intermediate-mass stars within 
$100\,{\rm pc}$ was initiated in 2007 in order to avoid the biases inherent in previous studies. This 
work yielded the magnetic properties of a large number of Ap/Bp stars in the sample; however, nearly 
half of the sample remained either unobserved or had relatively poor constraints on their field 
strengths and geometries. We have recently completed this survey using measurements obtained by ESPaDOnS 
and NARVAL. We discuss here some of the recent findings of this survey.
\end{abstract}

\section{Introduction}\label{sect:intro}

The generation and broader characteristics of magnetic fields of cool stars are reasonably well 
understood within the framework of stellar dynamo theory \citep[e.g.][]{Charbonneau2010}. In contrast, 
the origin of the magnetic fields of main sequence stars (MS) more massive than about $1.5\,M_\odot$ 
remains a profound mystery. It is now reasonably well established that all chemically peculiar Ap/Bp 
stars, corresponding to roughly $10\%$ of MS A- and B-type stars, host organized (primarily dipolar) 
magnetic fields with strengths 
$\lesssim30\,{\rm kG}$ \cite[e.g.][]{Wolff1968,Landstreet1982,Shorlin2002}. However, these results have 
typically been derived from magnitude-limited surveys or from surveys with relatively high detection 
thresholds and thus, are inherently biased.

In 2007, \citet{Auriere2007} attempted to explore the weak field regime of Ap/Bp stars by obtaining 
high-precision longitudinal field measurements of 28 objects with reportedly weak or otherwise poorly 
constrained field strengths. All of the observed Ap/Bp stars were found to exhibit dipolar field 
strengths of $B_{\rm d}\gtrsim300\,{\rm G}$ with the two weakest fields found to have 
$B_{\rm d}=100_{-100}^{+392}\,{\rm G}$ and $B_{\rm d}=229_{-76}^{+248}\,{\rm G}$. \citet{Auriere2007} 
hypothesized that there exists a critical field strength ($B_{\rm c}\approx300\,{\rm G}$), which 
corresponds to the minimum field strength that an Ap/Bp star must host in order to be invulnerable to a 
pinch-instability \citep{Tayler1973,Spruit2002}. More recently, a small number of A- and B-type stars 
hosting fields having $B_{\rm d}$ well below the proposed critical field strength have been identified 
in apparent contradiction to this explanation \citep[e.g.][]{Lignieres2009,Petit2011,Alecian2016}. 
While these recent discoveries call into question whether or not the so-called "magnetic desert" 
truly exists, the question remains why the vast majority of known Ap/Bp stars host fields with 
$B_{\rm d}\gtrsim300\,{\rm G}$.

In the following article, we present several results from a volume-limited spectropolarimetric survey of 
Ap/Bp stars located within $100\,{\rm pc}$. This survey has been carried out in order to constrain 
various fundamental and magnetic properties of this stellar population while attempting to minimize 
observational biases. This work was initiated by \citet{Power2007_MSc}, who obtained a large number of 
measurements using the now retired MuSiCoS instrument. These observations allowed for the magnetic 
parameters of approximately $50\,\%$ of the Ap/Bp stars in the sample to be derived. More recently, we 
have completed this survey primarily using Stokes $V$ measurements obtained with ESPaDOnS along with a 
small number obtained with NARVAL. In Section \ref{sect:sample}, we briefly describe the sample and 
present the derived incidence rate of Ap/Bp stars with respect to stellar mass. In Section 
\ref{sect:mag}, we present several results from our analysis of the Ap/Bp magnetic properties and, 
in Section \ref{sect:conc}, we state our conclusions.

\section{The Sample}\label{sect:sample}

\begin{figure}
  \centering
  \begin{subfigure}[b]{0.49\textwidth}
    \centering
    \centerline{\includegraphics[width=1.05\textwidth,clip=]{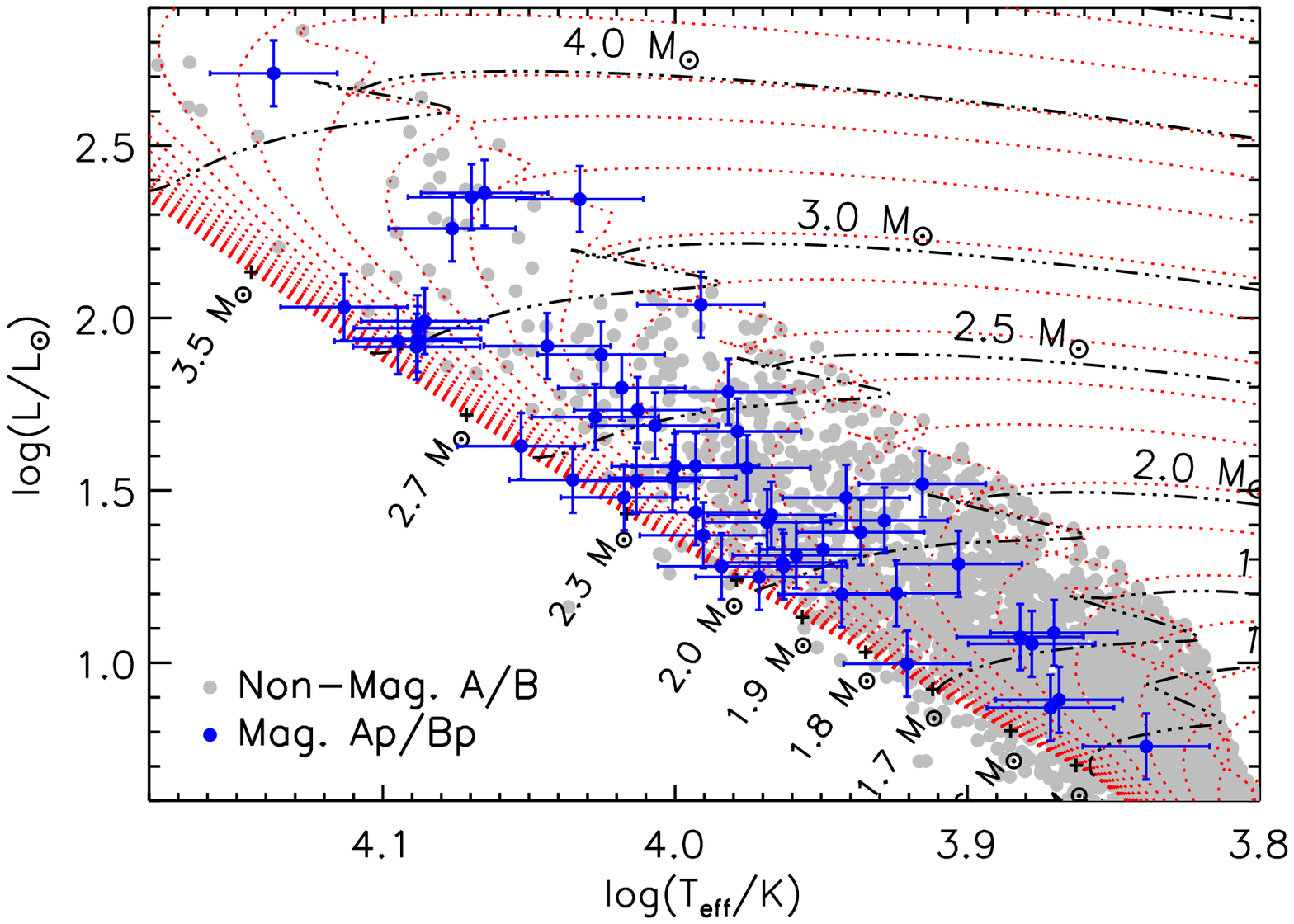}}
  \end{subfigure}
  ~
  \begin{subfigure}[b]{0.49\textwidth}
    \centering
    \centerline{\includegraphics[width=1.05\textwidth,clip=]{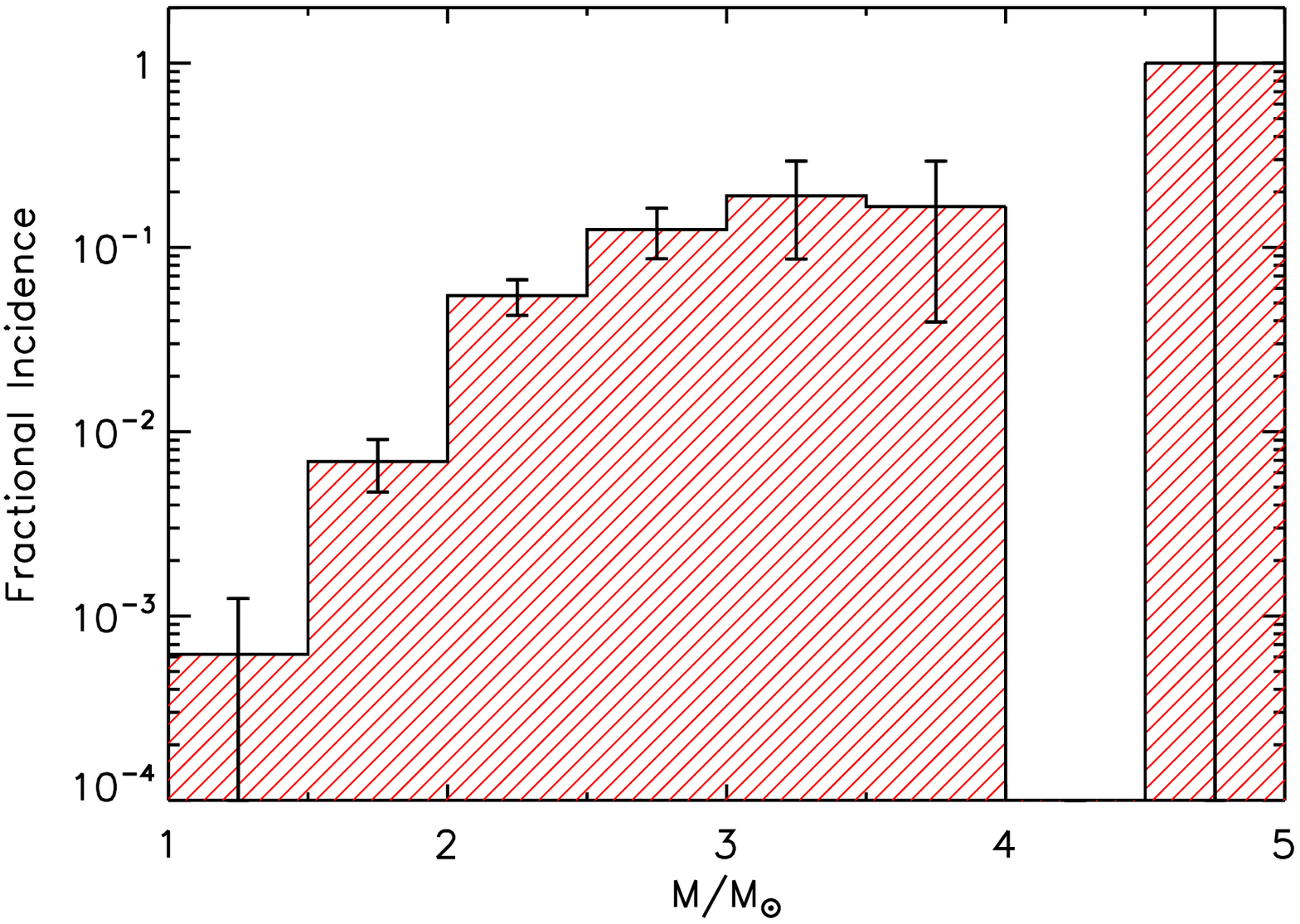}}
  \end{subfigure}
\caption{\emph{Left:} The Hertzsprung-Russell diagram for the volume-limited sample. Grey points 
correspond to the presumably non-magnetic MS stars while the blue points correspond to the known 
magnetic (Ap/Bp) MS stars. \emph{Right:} Incidence rate of Ap/Bp stars with respect to stellar mass. The 
large uncertainty and $100\,\%$ incidence rate associated with the highest mass bin 
($4.5\leq M/M_\odot<5$) is due to the fact that it contains only one star.}
\label{fig:fund}
\end{figure}

The full sample of intermediate-mass stars (magnetic and non-magnetic early-F, A-, and B-type stars) was 
compiled by identifying all objects within the Hipparcos catalogue having parallax 
angles $>10\,{\rm mas}$ ($d<100\,{\rm pc}$) \citep{VanLeeuwen2007}. This list was then cross-referenced 
with the Catalogue of Ap, HgMn and Am Stars \citep{Renson2009} in order to identify both known and 
candidate Ap/Bp stars. This yielded a total of 139 stars, which were classified as being unlikely, 
probably, or definitely Ap/Bp stars based on whether they had reported (1) photometric variability, (2) 
chemical peculiarities consistent with Ap/Bp stars \citep[identified either spectroscopically or through 
the use of photometric indices; i.e.][]{Paunzen2005}, or (3) magnetic detections. We obtained 327 Stokes 
$V$ observations of 65 stars using MuSiCoS (185 measurements), ESPaDOnS (114 measurements), and NARVAL 
(28 measurements). Based on our magnetic measurements, archived magentic measurements, and on published 
magnetic measurements, we conclude that the sample of early-F, A-, and B-type stars within a 
$100\,{\rm pc}$ volume contains 52 confirmed Ap/Bp stars and $\sim3\,700$ non-magnetic stars.

We derived each of the (presumably) non-magnetic stars' effective temperatures and luminosities by 
applying various photometric calibrations to a wide range of archived photometric measurements 
\citep[e.g.][]{Mermilliod1997a}. A more detailed analysis of the fundamental parameters was performed on 
the 52 Ap/Bp stars. This involved fitting both the available photometry and spectroscopy (including 
Balmer lines, which are relatively sensitive to $\log{g}$, and multiple $25-100\,{\rm \AA}$ width 
spectral regions containing He and metallic lines) to synthetic models. The synthetic SEDs were 
generated using the {\sc llmodels} code \citep{Shulyak2004} while accounting for flux abnormalities 
caused by chemical peculiarities and, in the case of strongly magnetic stars 
($B_{\rm d}\geq5\,{\rm kG}$), the anomalous Zeeman effect. All of the stars' stellar masses and ages 
were then estimated through comparisons with the evolutionary models of \citet{Ekstrom2012}. The total 
sample of MS stars is shown plotted on the Hertzsprung-Russell diagram in Fig. \ref{fig:fund} (left). 
The incidence rate of Ap/Bp stars as a function of stellar mass is shown in Fig. \ref{fig:fund} (right); 
it is evident that the incidence rate increases from $\lesssim1\,\%$ at $M\lesssim2\,M_\odot$ and 
plateaus at $\approx10\,\%$ for $M>2\,M_\odot$.

\section{Magnetic Properties}\label{sect:mag}

\begin{figure}
  \centering
  \begin{subfigure}[b]{0.49\textwidth}
    \centering
    \centerline{\includegraphics[width=1.05\textwidth,clip=]{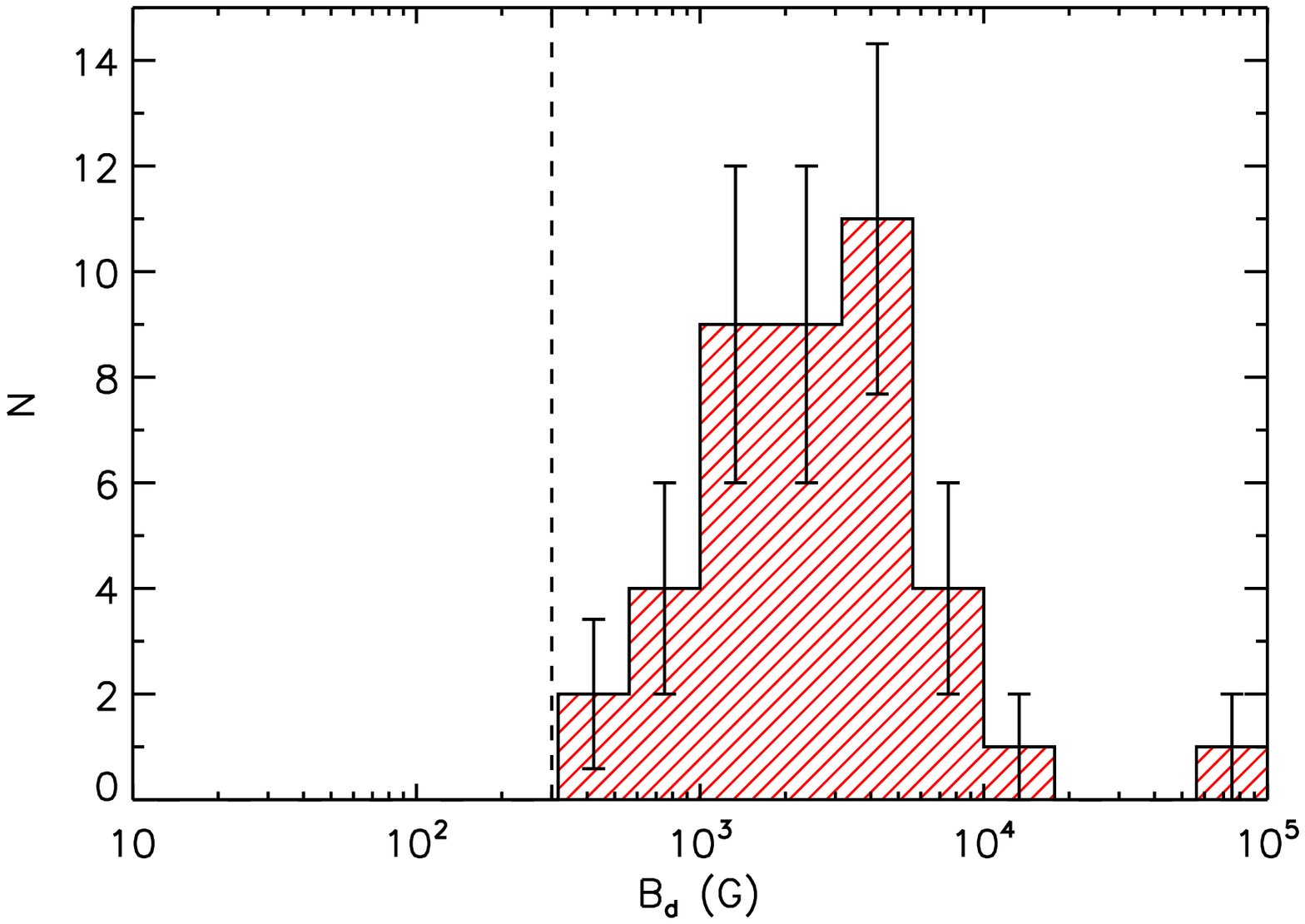}}
  \end{subfigure}
  ~
  \begin{subfigure}[b]{0.49\textwidth}
    \centering
    \centerline{\includegraphics[width=1.05\textwidth,clip=]{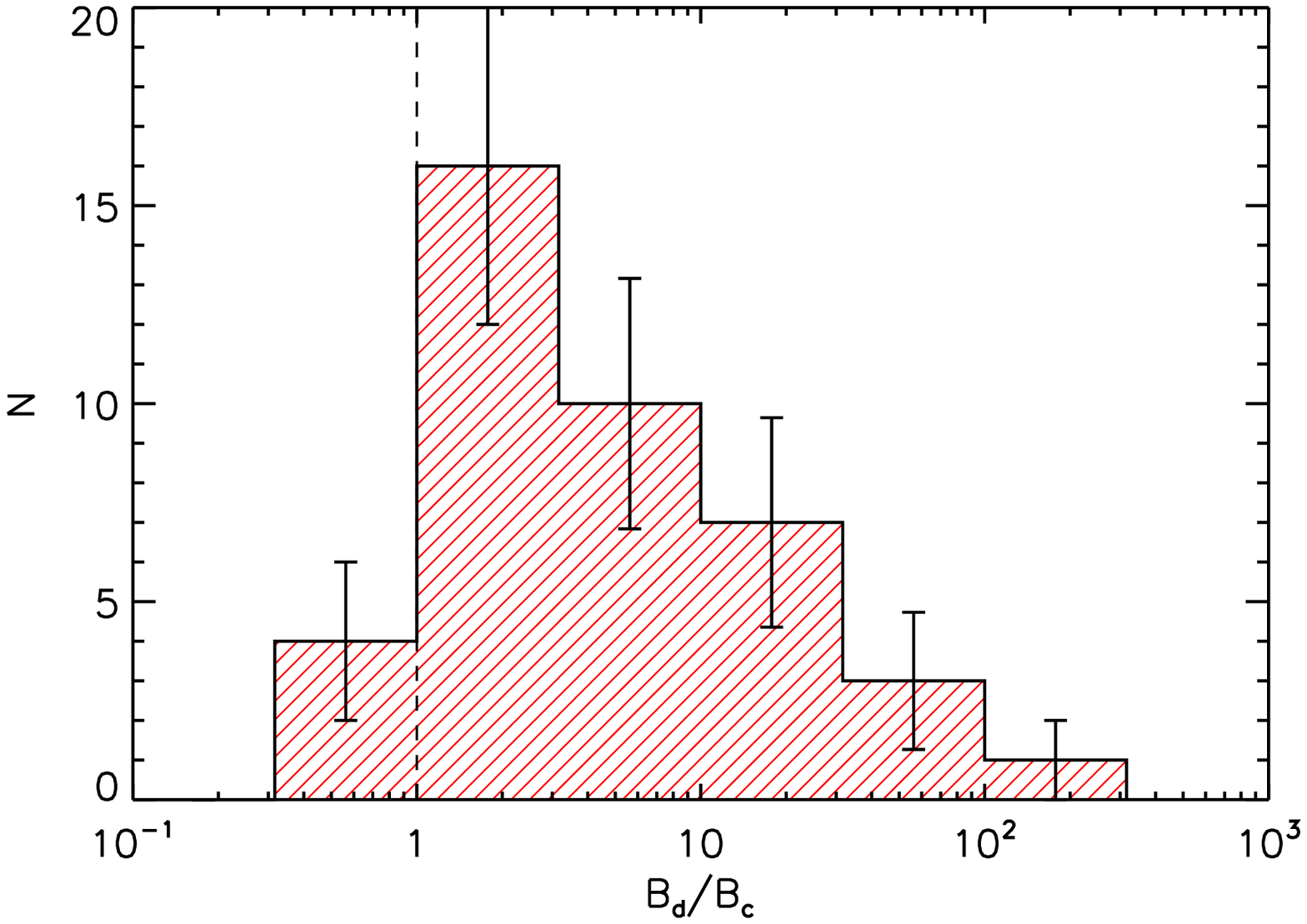}}
  \end{subfigure}
\caption{\emph{Left:} The distribution of the derived dipole magnetic field strengths. The 
vertical dashed line indicates the critical field strength ($B_{\rm c}=300\,{\rm G}$) derived 
by \citet{Auriere2007} for a typical Ap star. \emph{Right:} The ratio of each Ap/Bp star's dipole field 
strength to its critical field strength where $B_{\rm c}$ is calculated using Eqn. 8 of 
\citet{Auriere2007}.}
\label{fig:mag_dist}
\end{figure}

Both the dipole magnetic field strength and the obliquity angle $\beta$ (i.e. the angle between the 
dipole field's axis of symmetry and the star's rotational axis) can be derived if the inclination angle 
$i$ and the extrema of the longitudinal field ($B_{\rm z}$) are known \citep{Stibbs1950}. We attempted to 
derive these properties by obtaining approximately five $B_{\rm z}$ measurements per star such that the 
rotational periods could be roughly sampled. We adopted a target precision of 
$15\lesssim\delta B_{\rm z}\lesssim25\,{\rm G}$, which allowed for the detection of surface magnetic 
fields with $B_{\rm d}\gtrsim150\,{\rm G}$. The resulting distribution of $B_{\rm d}$ values is 
shown in Fig. \ref{fig:mag_dist} (left). The distribution peaks at approximately $2.5\,{\rm kG}$ and 
contains a minimum field strength of $340\,{\rm G}$ -- slightly above the estimated critical 
field limit of $B_{\rm c}=300\,{\rm G}$ calculated using Eqn. 8 of \citet{Auriere2007}. $B_{\rm c}$ 
depends on each stars' $T_{\rm eff}$, rotational period, and radius; therefore, we derived the ratio, 
$B_{\rm d}/B_{\rm c}$, for each star yielding the distribution shown in Fig. \ref{fig:mag_dist} (right). 
We find that all but four of the Ap/Bp stars in our sample exhibit $B_{\rm d}/B_{\rm c}>1$.

\begin{figure}
  \centering
  \begin{subfigure}[b]{0.49\textwidth}
    \centering
    \centerline{\includegraphics[width=1.05\textwidth,clip=]{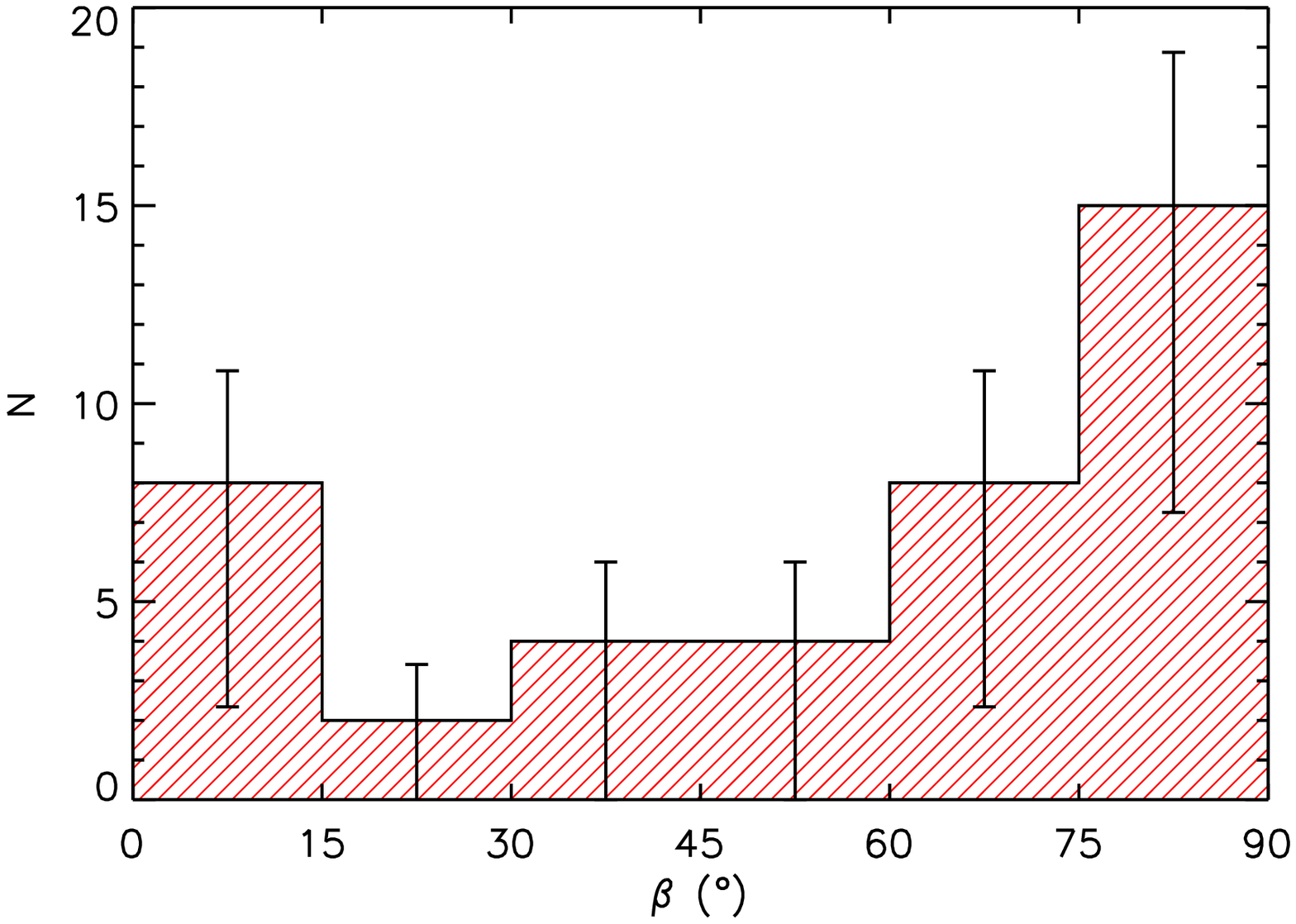}}
  \end{subfigure}
  ~
  \begin{subfigure}[b]{0.49\textwidth}
    \centering
    \centerline{\includegraphics[width=0.93\textwidth,clip=]{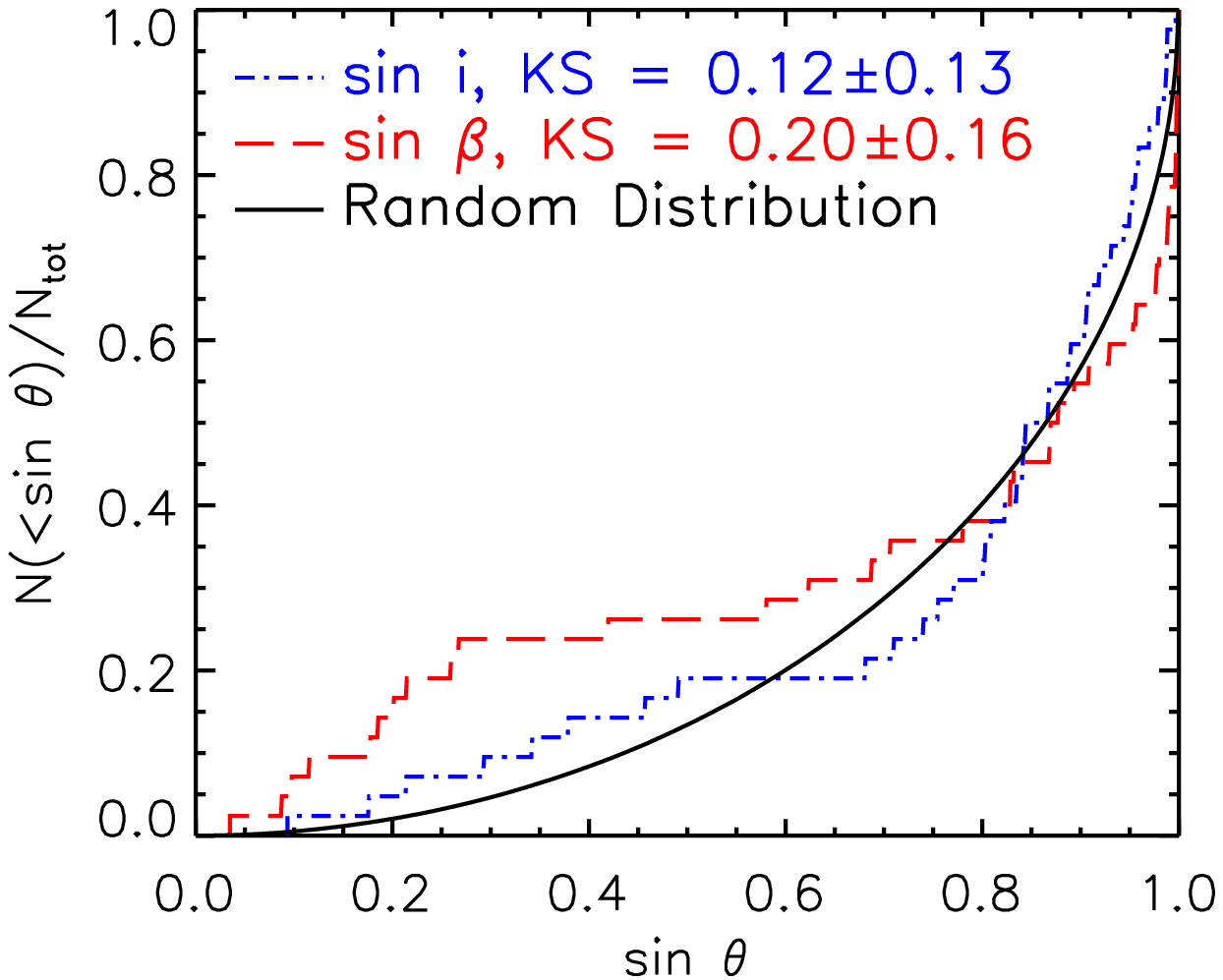}}
  \end{subfigure}
\caption{\emph{Left:} Distribution of derived obliquity angles. \emph{Right:} Cumulative distribution 
functions associated with the inclination angle (blue, dash-dotted) and obliquity angle (red, dashed). 
The solid black curve is generated from a distribution of randomly oriented axes. The $\sin{i}$ and 
$\sin{\beta}$ distributions are compared to the random distribution yielding Kolmogrov-Smirnov 
statistics of $0.12\pm0.13$ and $0.20\pm0.16$, respectively.}
\label{fig:inc_beta}
\end{figure}

Our analysis also yielded the obliquity angles associated with the dipolar field component of each 
Ap/Bp stars' magnetic field; this distribution is shown in Fig. \ref{fig:inc_beta} (left) while the 
cumulative distribution functions for both the derived $\beta$ and $i$ values are shown in Fig. 
\ref{fig:inc_beta} (right). We find that, while the $i$ distribution is consistent with a distribution 
of randomly oriented axes, the $\beta$ distribution exhibits a slight excess of instances in which 
$\beta\lesssim30^\circ$. Given the relatively small sample size and the difficulty of accurately 
constraining $\beta$, this result may be considered to be insignificant. This is supported by the 
derived 3 sigma uncertainty in the Kolmogrov-Smirnoff test statistic of $0.16$ obtained by employing 
bootstrap resampling.

\section{Conclusions}\label{sect:conc}

We have completed a volume-limited spectropolarimetric survey of Ap/Bp stars located within a distance 
of $100\,{\rm pc}$. We do not find any stars in our sample hosting magnetic fields with dipolar 
components less than $300\,{\rm G}$, which supports the existence of the so-called ''magnetic desert" 
first reported by \citet{Auriere2007}. In the preceding article, we have summarized several results 
from this survey; additional findings, including those obtained from a chemical abundance analysis of 
the sample of Ap/Bp stars, will be presented in a forthcoming publication.

\bibliography{jsikora}

\end{document}